# Response of Low Gain Avalanche Detector Prototypes to Gamma Radiation

**Martin Hoeferkamp[1], Alissa Howard[2], Gregor Kramberger[2], Sally Seidel[1*], Josef Sorenson[1], Adam Yanez[1]**

[1]Department of Physics and Astronomy, University of New Mexico, Albuquerque, NM 87106, USA

[2]Department of Experimental Particle Physics, Jozef Stefan Institute and Department of Physics, University of Ljubljana, Ljubljana, Slovenia

**\* Correspondence:**
Corresponding Author
seidel@unm.edu

**Abstract**

Motivated by the need for fast timing detectors to withstand up to 2 MGy of ionizing dose at the High Luminosity Large Hadron Collider, prototype low gain avalanche detectors (LGADs) have been fabricated in single pad configuration, 2x2 arrays, and related p-i-n diodes, and exposed to Co-60 sources for study.  Devices were fabricated with a range of dopant layer concentrations and, for the arrays, a variety of inter-pad distances and distances from the active area to the edge.  Measurements of capacitance versus voltage and leakage current versus voltage have been made to compare pre- and post-irradiation characteristics in gain layer depletion voltage, full bulk depletion voltage, and breakdown voltage.  Conclusions are drawn regarding the effects of the gammas both on surface and interface states and on their contribution to acceptor removal through non-ionizing energy loss from Compton electrons or photoelectrons.  Comparison of the performances of members of the set of devices can be used to optimize gain layer parameters.

## 1    Introduction

The Low Gain Avalanche Detector (LGAD) [1,2,3], based on the planar technology, produces a signal in response to generation of free carriers by a charged particle or high-energy photon, when operated depleted by a reverse bias.  An evolution of the avalanche photodiode (APD), the LGAD exhibits internal signal gain in the range of up to 100 that is proportional to the applied bias voltage.  The profile of the LGAD structure is $n^+/p/p^-/p^+$, where the p-implant below the highly doped $n^+$ cathode electrode is referred to as the multiplication implant, forming a gain layer, and has dopant concentration in the range a few times $10^{16}$ cm$^{-3}$ and depth 0.8 - 2.5 μm.  Detection of sub-nanosecond signals produced by minimum ionizing particles is possible with these devices [4].

LGADs are proposed for use in several experiments including upgrades to those at the Large Hadron Collider (LHC) [5,6].  A typical specification for their operation at the LHC, as components of the High-Granularity Timing Detector (HGTD) in ATLAS or the Endcap Timing Layer (ETL) of CMS, includes tolerance to 2 MGy of ionizing dose, which will accompany integrated hadronic fluence up to about $2.5 \times 10^{15}$ $n_{eq}$/cm$^2$ (this includes a safety factor of 1.5). Thin bulk is preferred, as the minimization of induced current variations due to Landau fluctuations will promote the best timing resolution [7].

An active area of research involves the problem of gain decrease as boron substitutional atoms deactivate in response to radiation damage; this is "acceptor removal" [8]. While the primary source of this problem is non-ionizing energy loss (NIEL) due to hadron radiation, a contribution also arises from the associated gammas, which produce point defects in the gain layer through the Compton (and to a lesser extent, photoelectric effect) electrons that they induce.

Gamma radiation motivates a second line of inquiry as well: characterization of oxide charge and interface traps in order to permit optimization of dimensions of the surface features, including inter-electrode separation and the distance between the active area and the edge. The goal is to maximize fill factor while ensuring against electrical breakdown under various operating scenarios.

## 2 Description of the Prototypes

Prototypes of three structures were produced by Hamamatsu Photonics K.K. (HPK) using epitaxial silicon grown on a Czochralski substrate; these are single LGADs, 2x2 ("quad") LGAD arrays, and associated p-i-n diodes. All have 50 μm active layer thickness, 200 μm total thickness, and a single guard ring. The n$^+$ electrode has dimensions 1.3 x 1.3 mm$^2$. The p-type gain layer is approximately 2.5 μm thick. All of the devices include under-bump metallization (UBM). Figure 1 (left) shows the surface of one such device, which also includes an opening for transient-current-technique (TCT) stimulation and a probe needle contact pad. The 2x2 arrays have the same features as the smaller devices but variations on inter-pad ("IP") separation (30, 40, 50, and 70 μm) as well as distance from the active area to the edge (300 and 500 μm). Figure 1 (right) shows a quad prototype. The p-i-n diodes have the same geometry as the LGADs but lack the gain layer. As they can tolerate high bias voltage while sustaining relatively little bulk damage, breakdown in the p-i-n's is indicative of breakdown in the bulk, typically at the guard ring where the field lines are focused. The LGADs and quads were produced with four different options on gain layer dopant concentration. Dopant concentrations of only a few percent difference have previously been shown to lead to very large differences in gain [7].

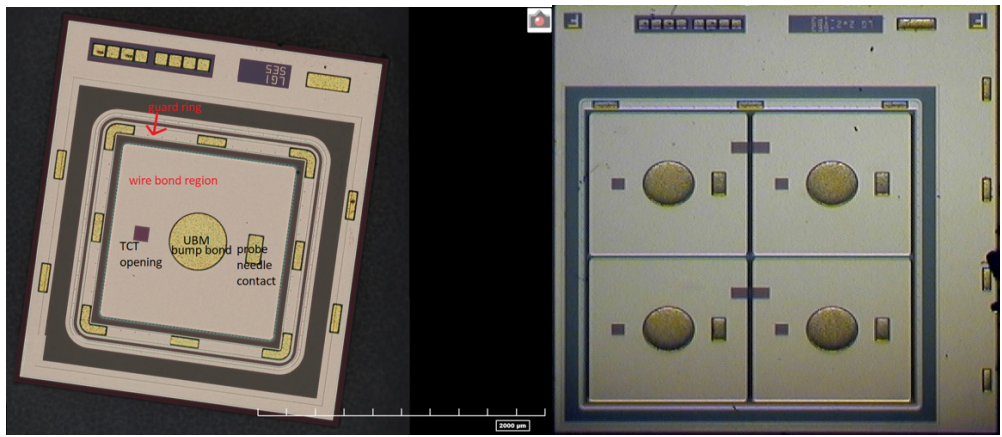

Figure 1. Photographs of (left) a prototype LGAD and (right) a quad sensor prototype.

## 3 Single LGAD Studies

### 3.1 Measurements

Prototypes were exposed to gammas at the Sandia National Laboratories Gamma Irradiation Facility for total ionizing doses in the range 0.1 to 2.2 MGy. Measurements of leakage current versus bias voltage ("IV") and capacitance C versus bias voltage ("CV") were carried out before and after the exposure. Figure 2 shows sample IV curves for a set of devices from Wafer 31; these represent the unirradiated characteristic as well as the response to doses of 0.1, 0.5, 1.0, and 2.2 MGy. The principal features of the curves are representative of all of the wafers studied, although differences were observed for the different gain layer concentrations, and they are discussed below. Noteworthy among the principal features are (1) the logarithmic rise in leakage current by approximately an order of magnitude during the initial application of bias, due to the surface component; (2) the knee at approximately 52 V, indicating the depletion of the gain layer; (3) increase of current (by approximately a factor of 5) which is moderate compared to the increase close to breakdown, which occurs above 160 V; this increase is an indication of the gain; (4) saturation of the current at the pre-gain layer depletion voltage at 0.1 MGy; and (5) increase of the breakdown voltage, $V_{bd}$, with dose, up to about 205 V for the 2.2 MGy sample.

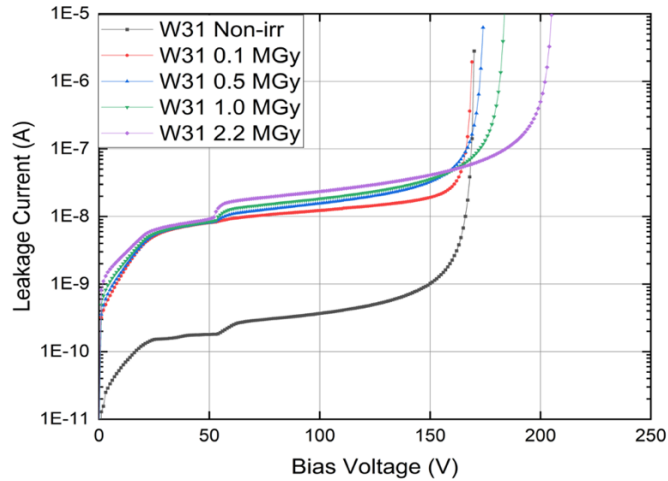

Figure 2. Sample current versus voltage characteristics for LGADs from Wafer 31, for various gamma doses.

Figure 3 shows an example curve of $1/C^2$ versus applied potential V, in this case for an unirradiated LGAD from Wafer 31; data recorded at a temperature of 20°C for applied signal frequency of 1 MHz are shown. The value of the frequency was varied between 1 kHz and 1 MHz, and there was no dependence of the results on the signal frequency provided by the HP4284A LCR meter. The two intercepts of the linear fits to the data in the three regions provide the gain layer depletion voltage $V_{gl}$ and the full bulk depletion voltage $V_{fd}$. Changes in gain layer depletion have previously been shown to correlate with charge collection performance [10].

### 3.2 Interpretation

Figures 4 and 5 show respectively the gain layer depletion voltage $V_{gl}$ and the difference between the full depletion voltage and $V_{gl}$, which should be proportional to the effective dopant concentration, as a function of total ionizing dose, for single LGAD devices representing all of the gain layer dopant concentrations in wafers with UBM. The data in Figure 4 are fit to the function $V_{gl} = V_{gl,0}\, e^{-c\phi}$, where $\phi$ is the total ionizing dose. Table 1 summarizes the extracted acceptor removal constant (c) values for each wafer, following exposure to 2.2 MGy. Also provided there are the pre-irradiation depletion voltages of the gain layers.

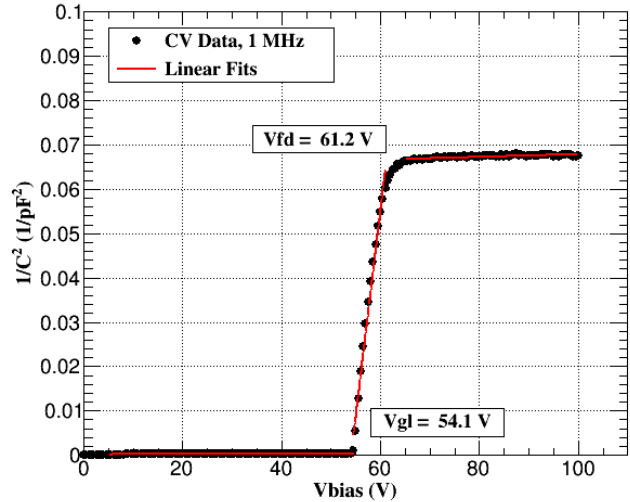

Figure 3. Sample capacitance characteristics versus bias voltage, for an unirradiated LGAD from Wafer 31.

The $V_{gl}$ is seen to be only slightly affected by even the highest dose. The full depletion voltage decreases slightly over the same range; this has been observed on epitaxial substrates also for proton-irradiated samples [11]. The observed increase in the breakdown voltage is validated by measurements on the p-i-n diodes (see below). The substantial rise of the surface current below depletion of the gain layer, i.e., below the point at which multiplication is possible, is not fully understood. The decrease of $V_{gl}$ in LGADs implies less multiplication. Once the LGAD is fully depleted, further increase of bias voltage adds to the field which can eventually reach breakdown level. The smaller the gain layer depletion voltage, the larger the breakdown voltage is for the device. For gain layer width approximately 2 microns, and active thickness 50 microns, every decrease of $V_{gl}$ by 1 V decreases the breakdown voltage by 25 V.

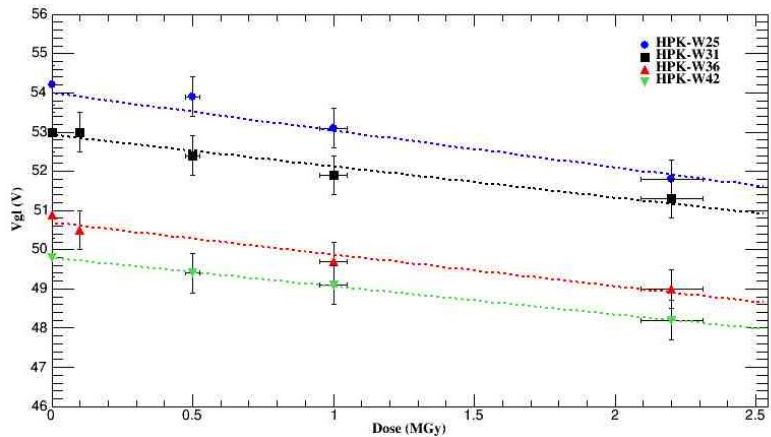

Figure 4. Gain layer depletion voltage as a function of dose, for LGADs from all four wafers. The fits to the function $V_{gl} = V_{gl,0}\, e^{-c\phi}$ are shown, and the resulting values of $c$ are reported in Table 1 as $c_\gamma$.

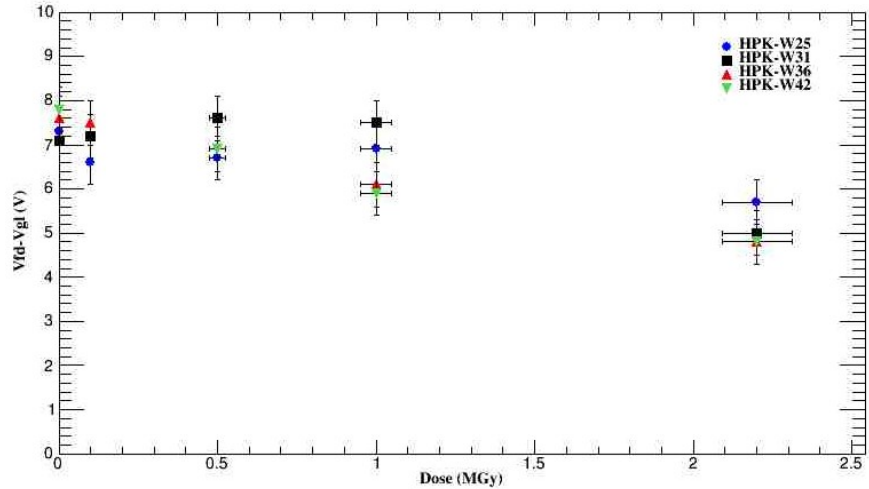

Figure 5. Full bulk depletion voltage minus gain layer depletion voltage, as a function of dose, for LGADs from all four wafers.

| Wafer # | $V_{gl,0}$ (V) | Inter-electrode separation in the quad sensors (μm) | $c_\gamma$ [x $10^{-8}$/Gy] after exposure to 2.2 MGy |
|---|---|---|---|
| 25 | 54 | 30, 40, 50, 70 | 1.79 ± 11.25% |
| 31 | 53 | 30, 40, 50, 70 | 1.53 ± 13.47% |
| 36 | 51 | 30 | 1.62 ± 15.17% |
| 42 | 50 | 30 | 1.47 ± 2.73% |

Table 1: Properties of the prototype wafers including the acceptor removal constants $c_\gamma$ of the gain layers, as obtained from a fit of the data in Figure 4 to the formula $V_{gl} = V_{gl,0}\ e^{-c\phi}$ after the gamma exposure reported here.

## 4 P-I-N Diode Measurements and Interpretation

Figure 6 shows example IV characteristics for p-i-n diodes from Wafer 25, for total ionizing dose from 0 to 2.2 MGy. In this case, post-irradiation breakdown voltages approaching 800 V are achieved, as irradiation-induced oxide charge moderates the electric field. If the LGADs' primary susceptibility to breakdown were also in the guard ring region, these high values would apply to them as well; however the LGADs break down at the electrode pads first. These outcomes follow closely the results of measurements made following neutron exposures, reported previously [12]. Between 700 and 800 V, the field in the bulk is sufficiently large that the device breaks down at its weakest point - regardless of whether it is a full LGAD or a p-i-n test structure. For the LGAD this breakdown is typically through the bulk, whereas for the p-i-n it is typically at the periphery.

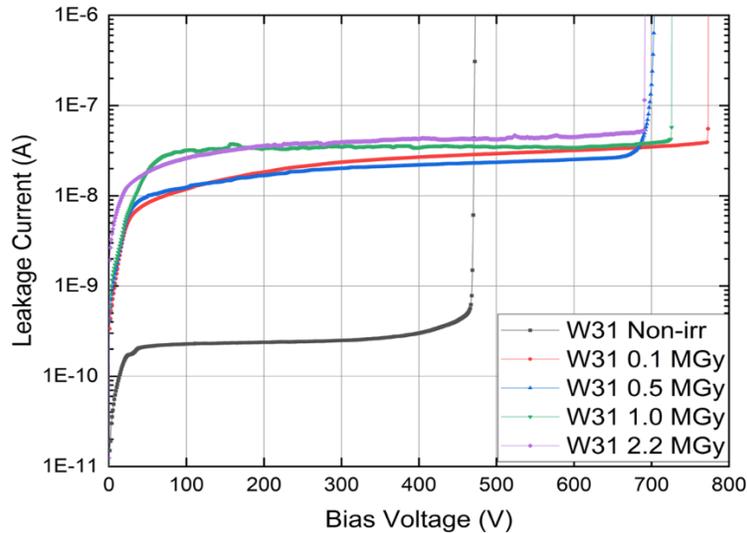

Figure 6. Sample current versus voltage characteristics for p-i-n diodes from Wafer 25, for several values of total ionizing dose.

## 5 Quad Sensor Measurements and Interpretation

If an electrode is floating, its potential is distributed to neighbors by punch-through. This process places a limit on the inter-electrode separation, for which the designer must anticipate the consequences in case a lost bump bond leads to breakdown at an electrode, which could then cascade to breakdown in neighbors.

An IV study involving the quad sensors was carried out to investigate the question of what minimal inter-electrode separation will reliably inhibit full punch-through. Bias is applied to the back side of the chip, and leakage current is measured with ground connected to the guard ring plus 0, 1, 2, 3, or all 4 pads. Figure 7 shows an example set of measurements of this type, for devices taken from Wafer 31, as a function of applied dose.

Figure 8 shows the punch-through voltage, as a function of dose, for all four inter-electrode separations. Punch-through between the guard ring and the pads occurs around 100 - 140 V prior to irradiation and decreases to nearly zero volts at 2.2 MGy, indicating the loss of resistivity in the region between the pads and the guard ring. At 2.2 MGy, all of the devices' IV curves are similar. In the case of the wafer with 30 μm inter-electrode separation, the breakdown voltage for measurement on the guard ring alone was observed to be substantially higher than the case in which the guard ring plus any non-zero number of pads are contacted. Among wafers with 30 μm inter-electrode separation but differing gains, the voltage at which punch-through occurs increases from approximately 85 V in Wafer 25 to approximately 100 V in Wafer 42.

**LGAD Response to Gamma Radiation**

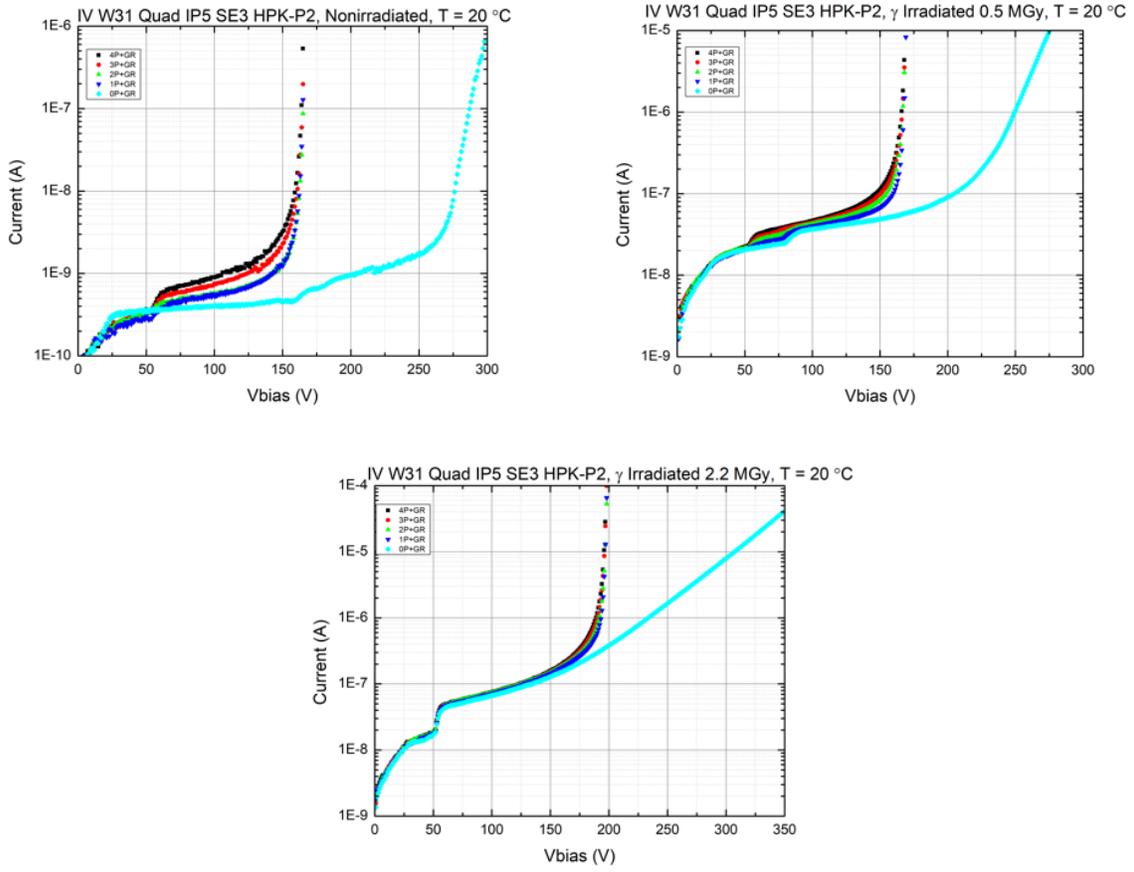

Figure 7. Leakage current versus bias voltage for quad sensors taken from Wafer 31, measured (upper left) prior to irradiation; (upper right) after exposure to 0.5 MGy; and (lower) after exposure to 2.2 MGy. In all cases the temperature during the measurement was 20°C. In each graph, measurements reflect the five modes in which the probes contact the guard ring plus *n* pads, where $n \in \{0,1,2,3,4\}$.

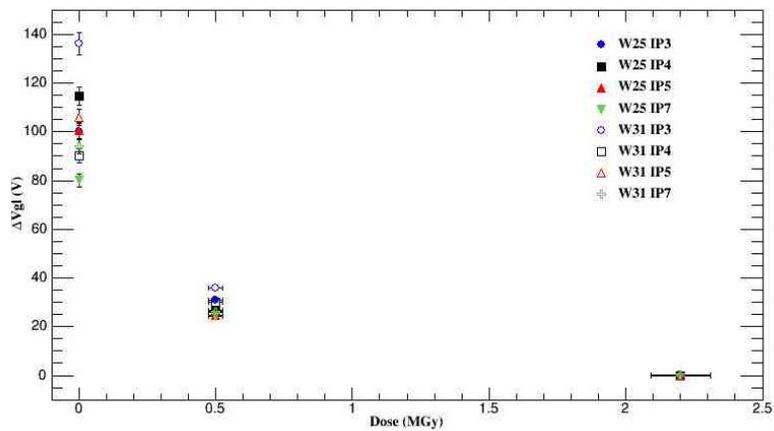

Figure 8. The difference between $V_{gl}$ for the electrodes connected directly to ground, and $V_{gl}$ for electrodes floating, as a function of dose and inter-electrode separation. This indicates the voltage needed for punch-through.

Figure 9 shows the leakage current versus applied bias voltage, for quad devices taken from all wafers (thus with four different initial values of the gain layer depletion voltage). These measurements were made after application of 0.5 MGy, for the measurement configuration indicated above, i.e., bias applied to the back side of the chip, and leakage current measured with ground connected to the guard ring plus 0, 1, 2, 3, or all 4 pads. Breakdown occurs at the same bias potential for measurements connecting the guard ring to any number of pads greater than zero. This indicates that the loss of a pad (e.g. disconnection of a bump) will present a danger of breakdown between that pad and its neighbors, for any of the inter-electrode separations (30 μm - 70 μm) reported here. It is interesting to note that by 2.2 MGy, while the IV curves are identical up to breakdown, the IV measured on the guard ring alone rises with a much slower characteristic.

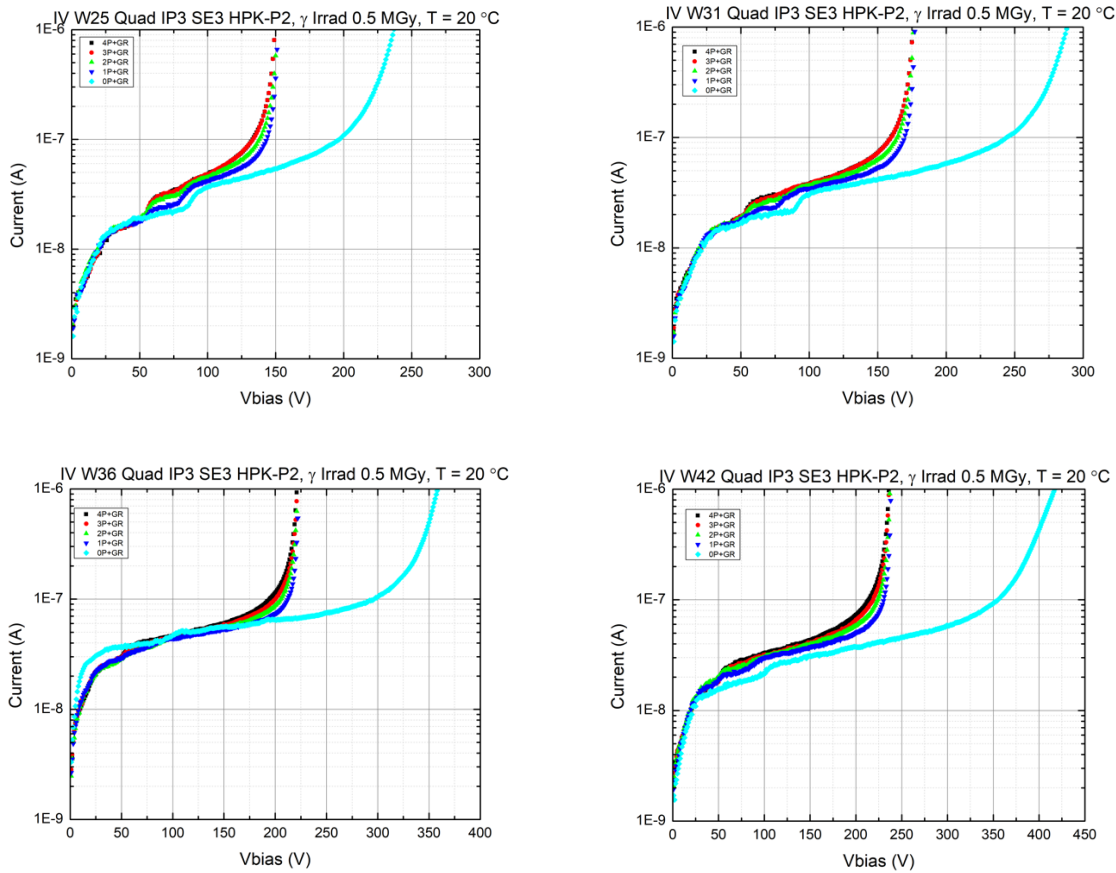

Figure 9. Leakage current versus applied bias voltage, for quad devices taken from all wafers (thus with four different initial values of the gain layer depletion voltage). These measurements were made at room temperature after application of 0.5 MGy, for the measurement configuration in which bias is applied to the back side of the chip, and leakage current is measured with ground connected to the guard ring plus 0, 1, 2, 3, or all 4 pads.

The resistance between electrodes in the quad sensors was also measured. On each quad sensor, one pad was biased to values in the range 0.5 V - 2.0 V relative to the remaining three grounded pads, and the current drawn on that biased pad from the others was recorded. The back side of the sensor was biased at -100 V with a separate source meter, and the guard ring was allowed to float. A fit to the slope of this IV characteristic yields the inverse of the resistance. Figure 10 shows the resistance values obtained in this way, for quad sensors representing the four inter-pad spacings, taken from wafers 25 and 31, for doses 0, 0.5, and 2.2 MGy. On all samples the resistance value is significantly greater than 1 GΩ prior to irradiation; it drops to values around 1 GΩ after application of 0.5 MGy; and reaches values in the range 10 - 100 MΩ after application of 2.2 MGy.

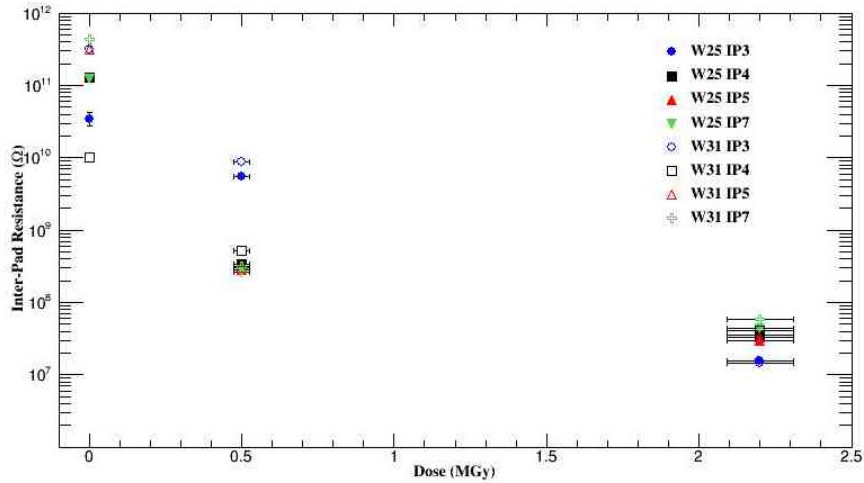

Figure 10. Inter-pad resistance measured on wafers 25 and 31, for all four inter-pad spacings, for doses 0, 0.5, and 2.2 MGy.

## 6      Uncertainties

The errors on the IV and CV measurements include statistical and systematic uncertainties. Each IV and CV data point shown on the graphs is the average of three to five measurements, and the standard deviation for each is found to be less than 2%. Systematic uncertainties include uncertainties associated with the setup configuration (typically 1.9%), the accuracy of the source and measurement instruments (±0.3%+100 fA for the Keithley 237; ±0.029%+300 pA for the Keithley 2410, and ±0.34% for the HP4284A); the precision of the measurement of the temperature (±0.5°C leads to uncertainty of ±1.82% on leakage current); and the data increment size of 1V. The uncertainties on $V_{gl}$, $V_{fd}$, and pad resistance derive from the quality of the linear fits and the bias voltage step size of ±1 V. Analyses of the measurements typically require linear fits, on which the uncertainty is typically a few percent.

## 7      Conclusions

The surface, gain layer, and bulk properties of the LGAD devices included in this study are found to change after gamma irradiation.

For the single LGAD devices and the p-i-n diodes, the surface component increases the total leakage current by more than an order of magnitude with only 0.1 MGy dose and saturates at about the same level with high gamma dose. The single LGADs have a much lower breakdown voltage than the p-i-n diodes, indicating that the LGAD breakdown occurs in the bulk at the electrode pad region. The gain layer and full depletion voltages both decrease by a small amount even at the highest dose, indicating some damage to the gain layer and bulk. The decrease in ($V_{fd}$ - $V_{gl}$) with dose implies a change in doping concentration in the gain layer and thus acceptor removal. The gamma radiation produced Compton electrons and photoelectrons that led to lattice point defects in the gain layer. The acceptor removal constant was characterized by fitting the data to a decaying exponential function (Figure 4), and the resulting values are shown in Table 1.

For the 2x2 quad LGAD devices, the punch-through between the guard ring and the pads for all inter-pad separations has been characterized to be over 100 V prior to irradiation; however after application of gamma irradiation it decreases for all devices studied and reaches nearly zero volts at the maximum 2.2 MGy dose. Pad-to-pad resistance after maximum dose is found to lie in the range 10-100 MΩ. Some variations in punch-through voltage and inter-pad resistance are observed for devices from different wafers and different doping concentrations.

# 6 Acknowledgments

This work was made possible by support from U.S. Department of Energy grant DE-SC0020255 and from ARRS and MIXS, Slovenia. It was also supported by the National Aeronautics and Space Administration (NASA) under Federal Award Number 80NSSC20M0034 (2020-RIG). The opportunity to use the Sandia Gamma Irradiation Facility, made possible by Dr. Maryla Wasiolek and Dr. Donald Hansen of Sandia National Laboratories, as well as by the University of New Mexico, is gratefully acknowledged. The support from Dr. Paulo Oemig of New Mexico State University and the encouragement of Dr. Jeremy Perkins and Dr. Regina Caputo, both of NASA/GSFC, is deeply appreciated.